\documentclass[review,12pt]{elsarticle}

\usepackage{lineno,hyperref}
\setpagewiselinenumbers
\modulolinenumbers[5]

\usepackage{graphicx}
\usepackage{amsmath}
\usepackage{amsfonts}
\usepackage{color,soul}
\usepackage{verbatim}

\usepackage{numcompress}

\biboptions{numbers,sort&compress}

\journal{CARBON}
\begin{document}

\begin{frontmatter}

\title{From Nanopores to Macropores: Fractal Morphology of Graphite}
\author[delft]{Zhou Zhou\corref{corresponding}}
\cortext[corresponding]{Corresponding author. Tel: +31(0)15 27 81612. E-mail address: Z.Zhou-1@tudelft.nl}
\author[delft]{Wim G Bouwman}
\author[delft]{Henk Schut}
\author[llb]{Sylvain Desert}
\author[llb]{Jacques Jestin}
\author[psi]{Stefan Hartmann}
\author[delft]{Catherine Pappas}

\address[delft]{Delft University of Technology, Faculty of Applied Sciences, Mekelweg 15, 2629JB Delft, The Netherlands}
\address[llb]{Laboratoire L\'{e}on Brillouin, CEA/CNRS, F-91191 Gif-sur-Yvette Cedex, France}
\address[psi]{Laboratory for Neutron Scattering, Paul Scherrer Institute, CH-5232 Villigen PSI, Switzerland}

\date{\today}
\begin{abstract}
We present a comprehensive structural characterization of two different highly pure nuclear graphites that compasses all relevant length scales from nanometers to sub-mm. This has been achieved by combining several experiments and neutron techniques: Small Angle Neutron Scattering (SANS), high-resolution Spin Echo SANS (SESANS) and neutron imaging. In this way it is possible to probe an extraordinary broad range of 6 orders of magnitude in length from microscopic to macroscopic length scales. The results reveal a fractal structure that extends from $ \sim 0.6 $ nm to 0.6 mm 
and has surface and mass fractal dimensions both very close to 2.5, a value found for percolating clusters and fractured ranked surfaces in 3D.
\end{abstract}

\end{frontmatter}
\linenumbers
\section{Introduction}
Graphite has been used as a neutron moderator in several types of nuclear reactors from the Chicago Pile 1 in 1942 to the more recent Very High Temperature Reactor (VHTR) and High Temperature Gas-cooled Reactors (HTGR). This synthetic polygranular material has a very high chemical purity and a complex microstructure, which affects the mechanical properties under extreme conditions and irradiation damage \cite{Haag2005}. \\
The crystallite structure and disorder of graphite at the atomic level can be investigated by neutron or X-ray diffraction \cite{Zhou2014}, and the microstructure by TEM, SEM or optical microscopy \cite{Wen2008,Jones2008,Kane2011,Karthik2012,Hacker2000}. 
On the other hand, the bulk mesoscopic structure of the pores can be explored by small angle scattering of X-rays (SAXS) or neutrons (SANS) \cite{thrower1996chemistry}. 
Very first SANS measurements on non-irradiated and irradiated nuclear graphites were performed in the 1960’s \cite{Martin1964} and 1970’s \cite{Martin1977Car,Martin1977JNM,Martin1978}. These results have been reinterpreted recently \cite{Mileeva2013} to
disclose a surface fractal structure from $ \sim0.2 $ to 300 nm, i.e.\ over three orders of magnitude in length. 
However, the graphite  inhomogeneities can be seen with an optical microscope or even with naked eye. Therefore an exploration over a larger range of length scales is necessary and for this purpose we have combined three neutron-based techniques: SANS, Spin Echo SANS (SESANS) and imaging to cover lengths from nm to mm. We investigated two different highly pure nuclear graphites, and the results show a fractal structure over an extraordinary large scale of lengths that spans  6 orders of magnitude and has fractal dimensions close to 2.5. This value is expected for several cases of percolating clusters \cite{Bradley1991,Saleur1987} and in the most general case of fractured ranked surfaces \cite{Schrenk2012} in three dimensions.
\section{Methods}
\subsection{Sample}
The samples were  disk-shape specimens with a thickness of 0.5 mm and a diameter of 16 mm cut from two types of nuclear graphite, designated as RID and PGA. The RID graphite was manufactured by Pechiney SA in the 1960's by baking a paste made of oil coke and pitch, graphitized by electrical heating, and was used at the research reactor of the Reactor Institute Delft. The PGA (Pile Grade A) graphite was manufactured by British Acheson Electrodes, Ltd. and Anglo Great Lakes, from needle shaped coke particles derived from the petroleum industry. It was used in the early gas-cooled reactors in UK and has been object of several investigations \cite{2006IAEAcharacterization,Jones2008,Hacker2000}. PGA was manufactured by extrusion, which leads to aligned coke particles along a direction $\hat{e}$ and thus to the anisotropic properties. For this reason we produced two series of samples: PGA1 cut perpendicular to $\hat{e}$, and thus isotropic; and PGA2 cut along $\hat{e}$ and thus anisotropic. For the sake of simplicity in the following we will focus on PGA1 and the results from PGA2 are averaged over the whole sample as for the other two samples. In this work, the supplement gives a detailed analysis of the results from the PGA2 sample and the anisotropy effect, which is weak but nevertheless visible.
\subsection{Neutron-matter interaction}  
Since neutrons are electrically neutral, they can penetrate into matter deeply, only interacting with the nucleii, and are valuable probes of the structure in the bulk. When a beam of thermal neutrons interacts with a sample it may be scattered or absorbed with a respective
probability that is given by the scattering lengths and the absorption cross sections of the specific elements and isotopes in the sample (for a comprehensive introduction to neutron scattering see \cite{Willis2009book}). These transmitted and scattered neutrons deliver structural information from the sample. In the following we will introduce the techniques used in our work.

\subsection{Small Angle Neutron Scattering (SANS)}
\begin{figure}	
	\centering
	\includegraphics[scale=0.5]{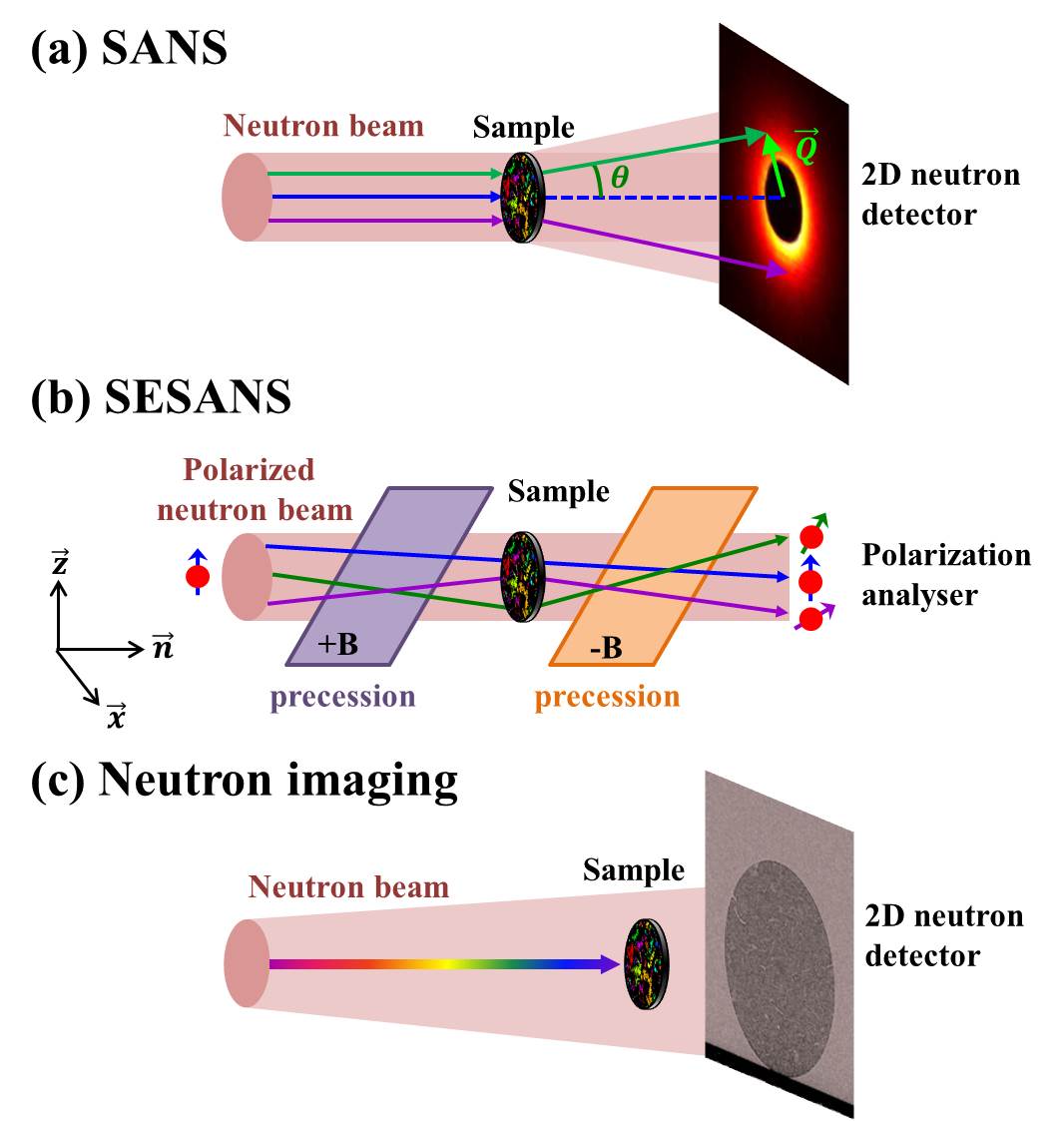}
	\caption{\label{fig:techniques}(a) Schematic diagram of a SANS experiment. A collimated monochromatic (wavelength $ \lambda $) neutron beam is scattered by structural inhomogeneities in the sample and recorded by a 2D position-sensitive neutron detector. The magnitude of the scattering vector is $ Q=| \protect\overrightarrow{Q} |=4\pi\sin(\theta/2)/\lambda $, with $ \lambda $ the neutron wavelength, and $ \theta $ the scattering angle.
	(b) Schematic diagram of a SESANS experiment. A polarized monochromatic neutron beam passes through the setup, where two magnetic fields in opposite direction before and after the sample induce Larmor preccessions. The initial polarization state is completely recovered for the transmitted beam as illustrated by the blue path, whereas scattering changes the final polarization state of the beam. 
	(c) Schematic diagram of a neutron imaging experiment. An incident polychromatic neutron beam is attenuated by the sample and the image recorded by a 2D camera-type neutron detector visualizes the structural inhomogeneities.}
\end{figure}

Fig. \ref{fig:techniques}a shows a schematic diagram of a SANS experiment. A neutron beam with a wavelength $ \lambda $ is scattered by the structural inhomogeneities (pores/carbon matrix in the case of graphite) and recorded by a 2D position-sensitive neutron detector. The transmitted beam is captured by a beam-stop made of neutron-absorbing material, illustrated by the black circular area in the resulting scattering pattern in Fig. \ref{fig:techniques}a.\\
SANS measures the intensity of scattered beam $ I(Q) $, i.e.\ the scattering cross section, as a function of the scattering vector $ \protect\overrightarrow{Q} $ that is related to the scattering angle $ \theta $
through $ Q=| \protect\overrightarrow{Q} |=4\pi\sin(\theta/2)/\lambda $. $ I(Q) $ can be factorized as \cite{Willis2009book}:
\begin{equation}
	\label{eq:SANS}
	I(Q)=\mathcal{B}\cdot P(Q)S(Q).
\end{equation}
Here $ \mathcal{B} $ is a pre-factor given by the neutron scattering length density contrast, in our case between the carbon matrix and the pores. $ P(Q) $ is the form factor characterizing the morphology/shape of the pores, and $ S(Q) $ is the structure factor corresponding to the correlations between pores.\\
In this work SANS measurements were performed at two instruments, the medium resolution PAXE and the high resolution TPA \cite{Desert2007} of the Laboratoire L\'{e}on Brillouin (LLB), CEA Saclay, France. On both instruments the neutron beam had a monochromatization of $\Delta\lambda/\lambda=10\%$. The experiments on TPA were done at $\lambda$=0.6 nm, covering the Q range $ 6\times10^{-3}\leq Q \leq 1\times10^{-1}$ nm$^{-1}$, and on PAXE at $\lambda$=0.37, 0.6, and 1.7 {nm}, respectively, covering the Q range $ 3\times10^{-2} \leq \; Q \; \leq \; 5 \;\mathrm{nm}^{-1}$. The PAXE data for $\lambda$=0.6 nm were brought to absolute units by normalization to the incident beam and were then used to normalise all other data using the large overlap in the Q-ranges illustrated by Fig. \ref{fig:SANS}a for the case of PGA1.
\subsection{Spin Echo Small Angle Neutron Scattering (SESANS)}
The basic principles of SESANS can be found in \cite{Rekveldt2005}, and Fig. \ref{fig:techniques}b represents schematically this technique. In contrast to conventional SANS, SESANS reaches high resolution accessing structural information on micrometres length scales by using a polarized neutron beam. Larmor precessions are induced in two regions with opposite magnetic fields, and if the setup is symmetric, the beam recovers its initial state, leading to a maximum spin echo polarization. Scattering from the sample breaks the symmetry and reduces the echo polarization. SESANS measures the spin echo polarization $ P_S(z) $ as a function of the spin echo length $ z $ \cite{Anderson2008}, the direction of which is determined by the geometry of the setup and it is always perpendicular to the propagation direction of the neutron beam $ \hat{n} $ (see Fig. \ref{fig:techniques}b). $ P_S(z) $ measures the projected scattering length density correlation function $ G^{'}(z) $:
\begin{equation}
\label{eq:SESANS}
P_S(z)=\exp[t\lambda^{2}(G^{'}(z)-G^{'}(0))/2\pi],
\end{equation}
and $ G^{'}(z) $ is the Hankel transformation of the SANS cross section:
\begin{equation}
\label{eq:Henkeltrans}
G^{'}(z)=\int_{0}^{\infty}J_0(Qz)I(Q)QdQ,
\end{equation}
where $ \lambda $ is the wavelength of the neutron beam, $ t $ is the sample thickness, and $ J_0 $ is a zeroth-order Bessel function of the first kind.

The high-resolution neutron Spin Echo SANS experiments were performed on the dedicated instrument of the Reactor Institute Delft \cite{Rekveldt2005} at $\lambda$=0.205 nm, with $\Delta\lambda/\lambda=5\%$ and covered length scales from 30 nm to 20 $ \mu $m.
\subsection{Neutron imaging}
Fig. \ref{fig:techniques}c shows a schematic view of a neutron imaging experiment. The incident polychromatic neutron beam passes through a sample. The intensity is attenuated due to absorption and/or scattering and is recorded by a 2D camera-type neutron detector. Since different phases in the sample have different attenuation coefficients, the resulting image visualizes the structural inhomogeneities of the sample. \\
Neutron imaging was performed on the cold neutron facility, ICON, of the Paul Scherrer Institute, Switzerland \cite{Kaestner2011}. The samples were placed as close as possible to the detector and a resolution of $ \sim $ 30 $ \mu $m was obtained. The transmission images were normalized to the empty beam and treated to obtain 8-bit digital images. These  were further analysed with the DIPimage Matlab toolbox (http://www.diplib.org/) as described in the supplement.
\section{Results}
\subsection{SANS}
\begin{figure}	
	\centering
	\includegraphics[scale=0.5]{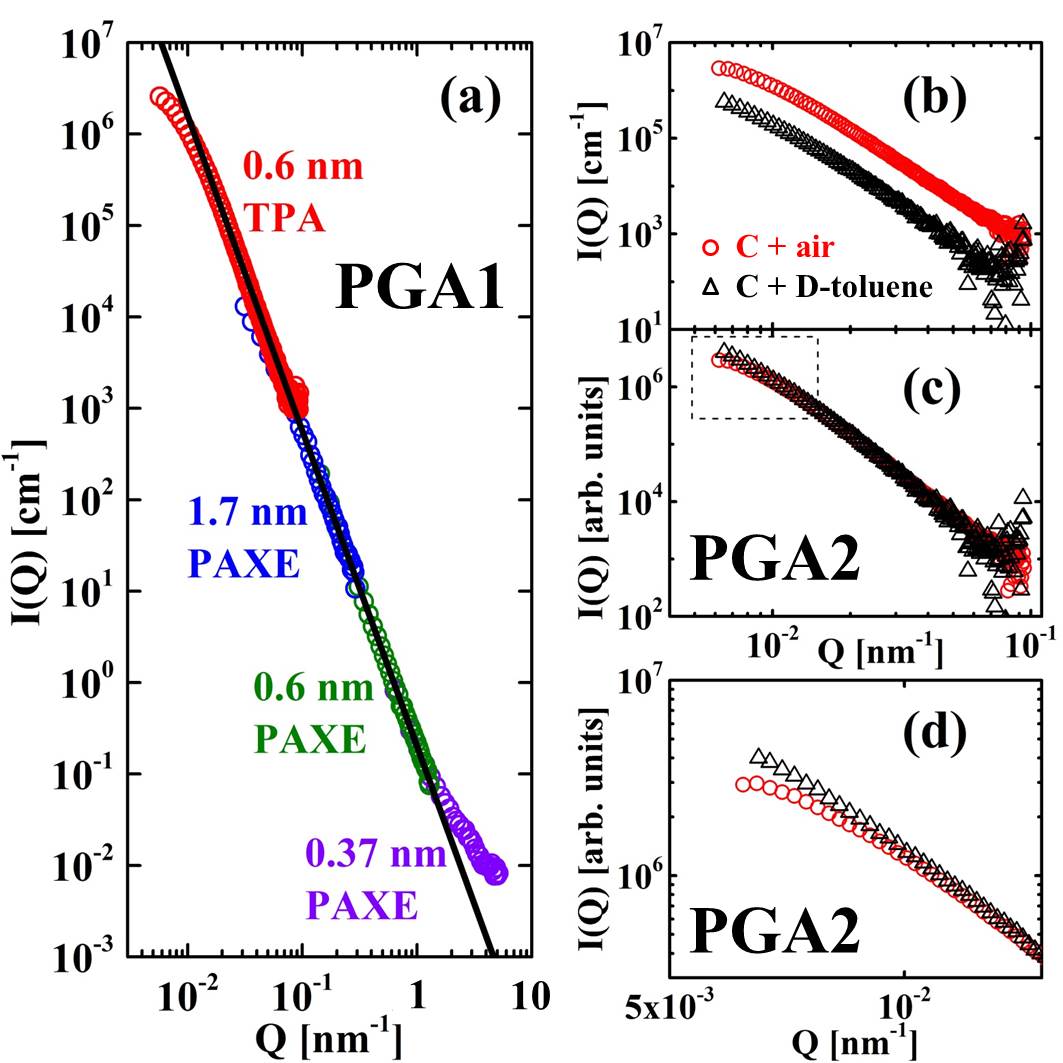}
	\caption{\label{fig:SANS}(a) Absolute scattering cross section $ I(Q) $ of PGA1 for the indicated different instrument and neutron wavelengths. The power law $ I(Q)\propto Q^{-\beta} $ with $ \beta = 3.45 $ is illustrated by the black line. (b)-(d) Contrast variation experiments to investigate the effect of multiple scattering: (b) $ I(Q) $ of PGA2 in air (red circles) and of the same sample embedded in deuterated toluene (black triangles). The data are vertically shifted to overlap in (c). A slight difference is found at the very low-Q range, which is enlarged in (d).}
\end{figure}
The absolute scattering cross section of PGA1, shown in Fig. \ref{fig:SANS}a, follows a power law $ I(Q)\propto Q^{-\beta} $ with $ \beta = 3.45\pm 0.01$. Similar results have been obtained for the other samples and the values of $\beta$ are given in Table \ref{tab:power}. Deviations at high-Q's are mainly due to (spin-incoherent) background, whereas at low Q's to multiple scattering. Graphite is a strong neutron scatterer, a property used for nuclear applications, and multiple scattering is not negligible even for the thin samples used in this study, as illustrated by the transmission values given in the supplement. For this reason SANS was also measured by embedding one sample (PGA2) in deuterated toluene, which reduces the scattering contrast. As a consequence  the transmission increased and the scattered neutron intensity decreased by almost an order of magnitude as shown in Fig. \ref{fig:SANS}b. However,  the shape stays unchanged with the exception of the very low-Q range illustrated by Fig. \ref{fig:SANS}c, where the two curves are shifted on the log-log scale. In the enlarged view of Fig. \ref{fig:SANS}d, it is clear that the bending characteristic for multiple scattering \cite{Schelten1980} is less pronounced for the sample in deuterated toluene. Thus multiple scattering has an influence on the SANS patterns but does not affect the power law, which  is the same for both cases and  reflects the genuine structural properties of the sample. Similar behaviour has been found in expanded graphite \cite{Balima2013} and sedimentary rocks samples \cite{Radlinski1999}.
\begin{figure}
	\centering
	\includegraphics[scale=0.5]{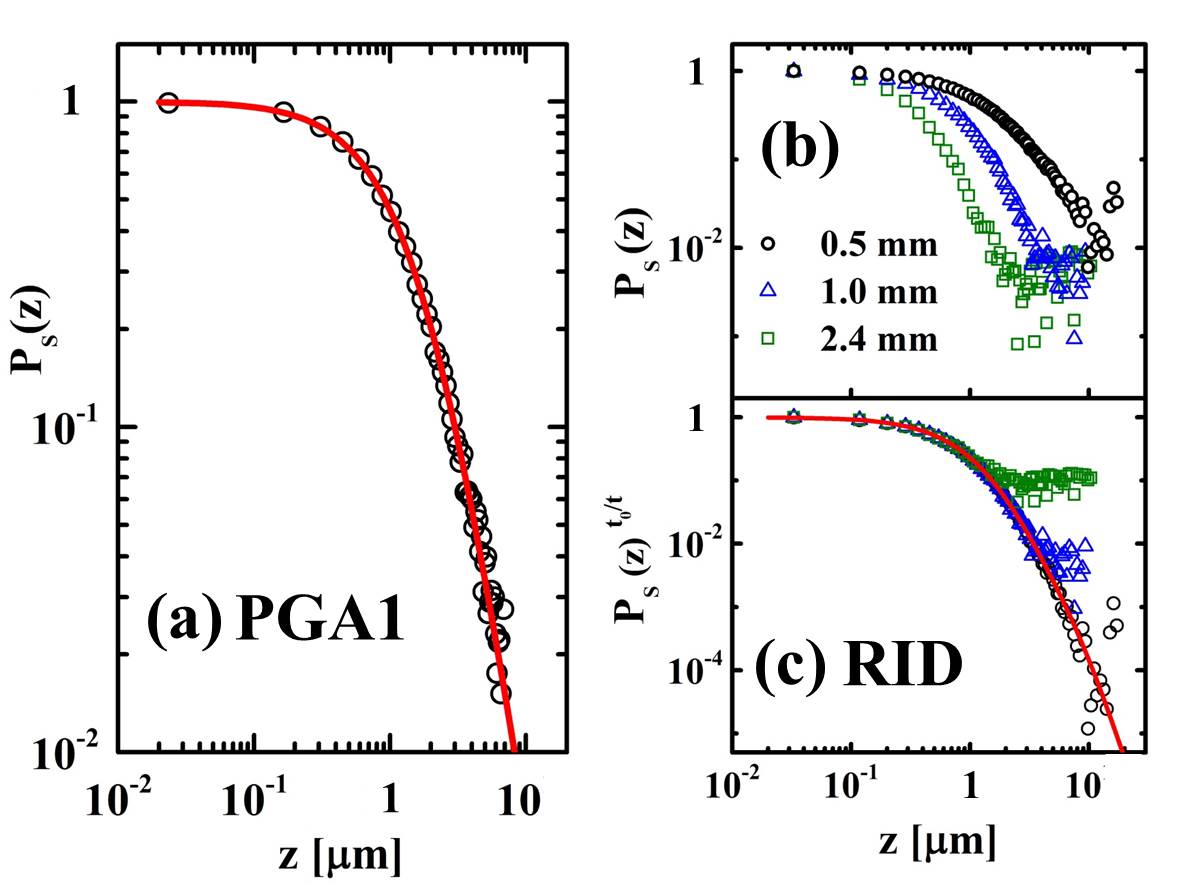}
	\caption{\label{fig:SESANS}Normalized SESANS polarization $P_S(z)$ of (a) PGA1 and (b) RID graphite samples for different sample thicknesses (0.5, 1.0 and 2.4 mm) as a function of the spin echo length $z$. This effect, which due to the scattering power of the sample and thus to multiple scattering, is taken into account in (c), where $ P_S(t)^{t_0/t} $ is plotted versus $ z $, for $ t_0=1 $ mm revealing the same generic curve for all samples. The solid lines in (a) and (b) correspond to the fitting curves from Hankel transform of Eq. \ref{eq:SANS} and \ref{eq:fractalmodel} with the parameters of Table \ref{tab:param}.}
\end{figure}
\subsection{SESANS}
Fig. \ref{fig:SESANS}a shows the SESANS pattern of PGA1. As already mentioned, SESANS measures the Hankel transformation of $ I(Q) $ in the form of normalised (to the direct beam) SESANS polarization $P_S$, which is a function of the spin echo length $z$ \cite{Anderson2008}. It probes length ranges from $\sim$ 30 nm to $\sim$ 20 $\mu$m, and thus "sees'' very large objects, which scatter a large fraction of the incoming beam. This is also the case in Fig. \ref{fig:SESANS}a, where even for the 0.5 mm PGA1 sample at large spin echo lengths $z$, the entire beam is scattered and $P_S (z>10 \, \mu m)\rightarrow 0$. However, even in this extreme case multiple scattering does not alter the results. This effect can indeed be accounted in a way similar to the Beer-Lambert's law in optics and if $ P_S(t_0) $ is the SESANS signal for a reference sample thickness $ t_0 $, the SESANS signal for any other thickness $ t $ is given by: $P_S(t)^{t_0/t}=P_S(t_0)$  \cite{Rekveldt2003}. This property was tested on three RID  samples with thicknesses of 0.5, 1 and 2.4 mm respectively. As shown in Fig. \ref{fig:SESANS}b, $ P_S (z) $ decreases much faster for the thicker samples. However, when plotting $ P_S(t)^{t_0/t} $, with $ t_0 $=1 mm, all data follow on the same generic curve shown in Fig. \ref{fig:SESANS}c, with deviations when $P_S$ is smaller than the experimental error of $ 10^{-2} $. More importantly, these results show that the porous structure extends up to macroscopic length scales, which can be investigated by direct imaging.
\subsection{Neutron imaging}
\begin{figure}
	\centering
	\includegraphics[scale=0.5]{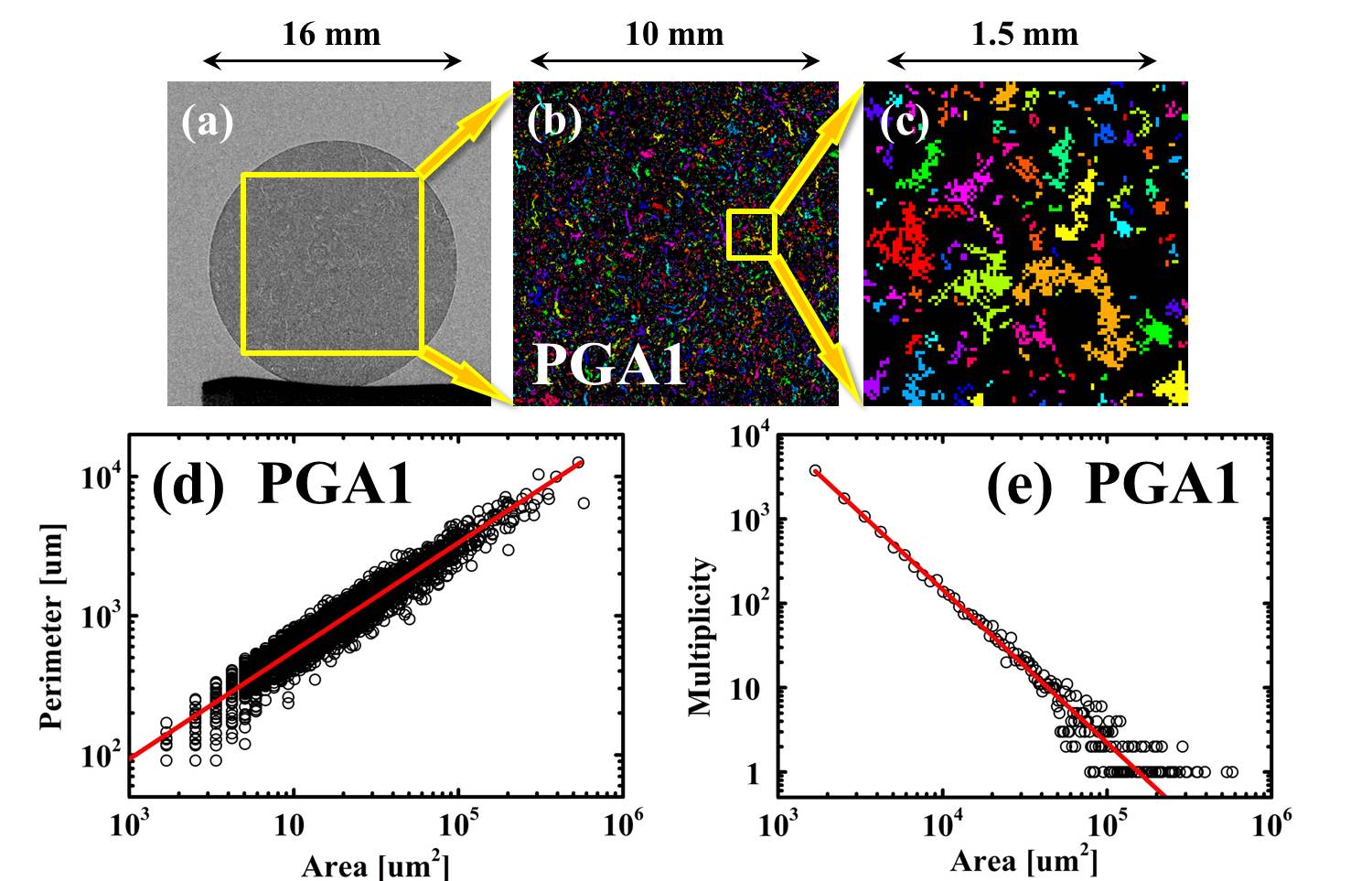}
	\caption{\label{fig:Imaging1}: Neutron transmission image of the PGA1 sample (a),  within the selected yellow square, the pores are labelled with different colours (b), and a selected region is enlarged in (c). The analysis of the shapes of the pores leads to the Perimeter-Area plot of (d). The red line corresponds to a power law $ P\propto A^\gamma $ with $ \gamma=0.77 $. The pore size distribution is given in (e), which shows a power law $ M(A)\propto A^\tau $ with $ \tau=-1.82 $ illustrated by the red line.}
\end{figure}
Fig. \ref{fig:Imaging1}a shows the neutron transmission image of the PGA1 sample (the images for the other samples are in the supplement). The treatment of this image (see supplement) leads to a binary image, where the pores are distinguishable from the carbon matrix. Consequently the pores within the selected yellow square (of size $ 800 \times 800 $ pixel) are labelled with different colours according to their sizes and shapes. This leads to Fig. \ref{fig:Imaging1}b, and a selected region enlarged in (c) reveals a strongly ramified, fractal structure. This structure may be brought in relation with the reported micropores \cite{Wen2008,Jones2008,Kane2011,Hacker2000}, that result from calcination and gas evolution during the manufacturing process and have sizes from several micrometres to hundreds of micrometres without preferred orientations. The binary images also lead to an estimation of the porosity $ \phi_{\text{image}} $ , which is about 82-85\% of the porosity $ \phi_D $ determined from density measurements, considering 2.25 g/cm$ ^3 $ as the density of single crystalline graphite. Thus pores smaller than the 30 $ \mu $m (resolution limit of the images) contribute only by 15-18\% to the total porosity. 

The plot of the pore perimeter $ P $ against area $ A $ for PGA1 (the plots for other samples are given in the supplement), shown in Fig. \ref{fig:Imaging1}d, reveals a power law $ P\propto A^\gamma $ with $ \gamma=0.77\pm0.01 $ (red line in the figure). This value is higher than 0.5, the value expected for Euclidian geometry, and is thus a signature of fractality. We note that in this figure the points have different multiplicities. 
The pore size multiplicity $ M(A) $ is plotted against $ A $ in Fig. \ref{fig:Imaging1}e, where $ M(A)\propto A^\tau $ with $ \tau=-1.82\pm0.03 $, illustrated by the red line in the figure, and the values of $ \tau $ for other samples are listed in Table \ref{tab:power}. Thus the total volume of pores $ V(A) $ with specific area size $ A $ is given by: $ V(A)\propto M(A)\times A^{3/2} \propto A^\tau\times A^{3/2}=A^{\tau+3/2}\sim A^{-0.3} $. The low value of the exponent indicates that the pores with different sizes have almost the same total volume.

\begin{figure}
	\centering
	\includegraphics[scale=0.5]{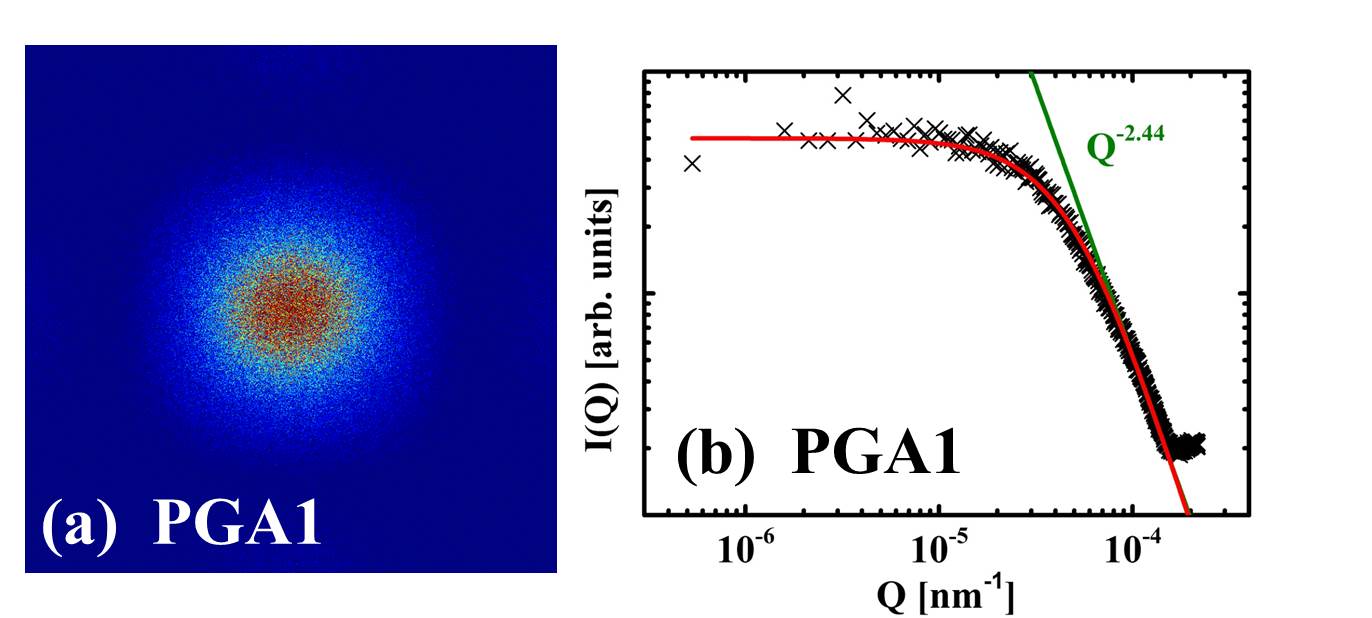}
	\caption{\label{fig:Imaging2}:The 2D scattering pattern of PGA1 obtained by 2D-Fourier transform on the selected region illustrated by the yellow square in Fig.\ref{fig:Imaging1}a. This leads to the $ I(Q) $ curve of (b). The power law  at $ Q\geq 1.5\times 10^{-4} $ nm$ ^{-1} $, $ I(Q)\propto Q^{-2.44} $ illustrated by the green line, has an exponent smaller than 3 indicating a mass fractal. The red line in (b) corresponds to Eq. \ref{eq:SANS} and \ref{eq:fractalmodel} with parameters of Table \ref{tab:param}.}
\end{figure}
A quantitative analysis of the transmission images involves a 2D-Fourier transform of the region inside the yellow square of Fig. \ref{fig:Imaging1}a, which leads to the 2D scattering pattern of Fig. \ref{fig:Imaging2}a. By radially averaging this intensity, following the same procedure as for the SANS data, the scattering curve of Fig. \ref{fig:Imaging2}b is obtained, with Q calculated as given in the supplement. The resulting $ I(Q) $ levels off at low-Q region and crosses over to a power law $ I(Q) \propto Q^{-2.44\pm0.09} $ at higher $Q$'s.
\section{Discussion}
\begin{figure}
	\centering
	\includegraphics[scale=0.5]{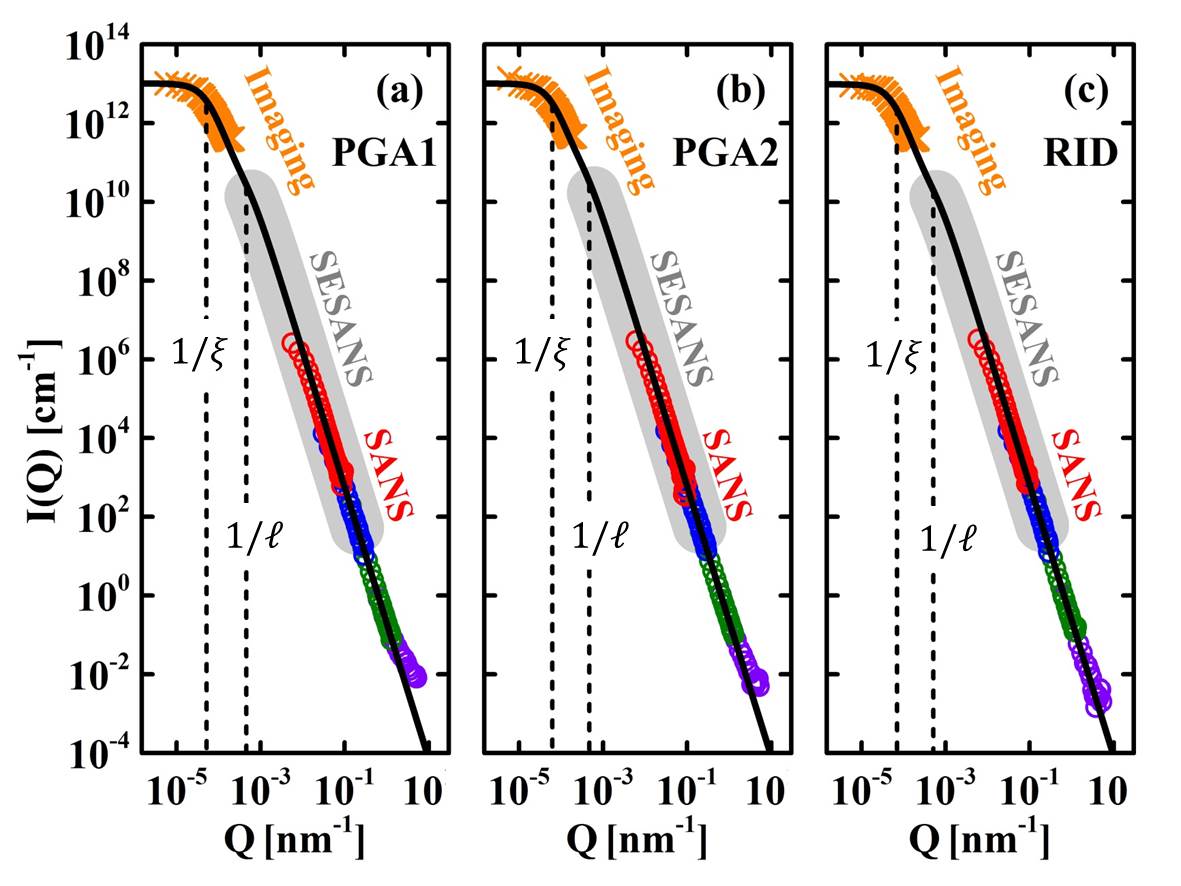}
	\caption{\label{fig:Global}Scattering curves for (a) PGA1 (b) PGA2 and (c) RID over for the whole Q range investigated. The circles represent the SANS data; the orange crosses are from the Fourier transformation of the neutron images; and the gray shaded zone represent the Q-range covered by SESANS. The black lines correspond to the best fit of the Eq. \ref{eq:fractalmodel}  with the parameters of Table \ref{tab:param}. The dotted lines correspond to $Q=1/\xi$ and $1/\ell$ respectively.}
\end{figure}

These results disclose power laws, which are characteristics of fractal topology: fractal surfaces (of pores or particles) with a surface fractal dimension $ D_s $ or mass fractals with a mass fractal dimension $ D_m $ \cite{Mildner1986, Schmidt1991}. The most direct signature of  fractality is  the power law of the SANS intensity shown in Fig. \ref{fig:SANS}. It is indeed expected that \cite{Bale1984,Teixeira1988}:
\begin{align}
	\label{eq:surfpower}
	\text{for a surface fractal:}\quad I(Q) &\propto Q^{-(6-D_s)},\\
	\label{eq:masspower}
	\text{for a mass fractal:}\quad I(Q) &\propto Q^{-D_m},
\end{align}
where the values of both $ D_m $ and $ D_s $ are smaller than 3, the dimensionality of the Euclidian space. Expressing the power laws as $ I(Q)\propto Q^{-\beta} $, $ \beta>3 $ corresponds to a surface fractal and $ \beta<3 $ to a mass fractal. All graphites showed $ \beta>3 $ leading to the surface fractal dimensions listed in Table \ref{tab:power}.\\
\begin{table}
	\centering
	\caption{\label{tab:power}Power law exponents ($ \beta $, $ \tau $), fractal dimensions ($ D_s $, $ D_m $) and porosity ($ \phi_D $ and $ \phi_{\text{image}} $) derived from SANS and imaging.}
	\resizebox{\textwidth}{!}{
		\begin{tabular}{cccccccc}
			\hline \hline
			&\multicolumn{2}{c}{SANS}&\multicolumn{3}{c}{Imaging}&\multicolumn{2}{c}{Porosity}\\
			Sample&$ \beta $&$ D_s $&$ Ds $&$ D_m $&$ \tau $&$ \phi_D $&$ \phi_{\text{image}} $\\ \hline
			PGA1 & 3.45 (1) & 2.55 (1) & 2.54 (2) & 2.44 (9) & -1.82 (3) & 22.7(5)\% & 18.6\% \\
			PGA2 & 3.43 (1) & 2.57 (1) & 2.50 (2) & 2.54 (9) & -1.80 (4) & 22.7(5)\% & 17.8\% \\
			RID  & 3.45 (1) & 2.55 (1) & 2.56 (2) & 2.45 (9) & -1.74 (3) & 27.1(5)\% & 23.0\% \\ \hline \hline& & & 
		\end{tabular}}
	\end{table}
Additional confirmation of these results comes from the Perimeter-Area power law $P\propto A^\gamma$ of Fig. \ref{fig:Imaging1}d, where the exponent $\gamma$ is directly related to the surface fractal dimension $D_s$ through $\gamma=(D_s-1)/2$  \cite{Madelbrot1983,Pentland1984,Beech1992}, leading to  $ D_s=2.54\pm0.02 $ for PGA1. Similar results were obtained for the samples and all deduced values of $ D_s $ are listed in Table \ref{tab:power} and are consistent with those obtained from SANS. Therefore imaging and SANS "see" the same surface self-similar fractal structure  although they probe length scales more than 3 orders of magnitude apart.\\
On the other hand, the power law of $I(Q)$  derived from the Fourier transformed images leads to an exponent  significantly smaller than 3 (Fig. \ref{fig:Imaging2}b), which is in line with a mass fractal and not surprising at this very low Q-limit. The deduced values of $ D_m $ are given in Table \ref{tab:power}, and for all samples $ D_m \approx D_s $.\\
The experimental results indicate  that the pores of the graphite samples have a fractal (i.e. rough) surface while their assembly forms a mass fractal. To be specific, one can consider that the pore structure consists of many pore-building blocks, which characterize the network of the pore clusters with a mass fractal property at the length scale above the size of the primary block. Moreover, the pore building block itself is bounded by a rough surface with fractal morphology. In this case the scattering function reflects both mass and surface fractal properties and Eq. \ref{eq:SANS} can be factorized as \cite{Teixeira1988,Wang2013}:
\begin{subequations}
	\label{eq:fractalmodel}
	\begin{align}
	\label{eq:prefactor}
	\mathcal{B} &=\phi_D \, \Delta \rho^{2} \,V_p=\phi_D \, \Delta\rho^{2} \,4\pi \, \ell^{3}/3 \\
	P(Q) &=(1+Q^{2}\ell^{2})^{(D_s-6)/2} \\
	S(Q)&=1+\frac{D_m \, \Gamma (D_m-1)}{(2Q\ell)^{D_m}}(1+\frac{1}{(Q\xi)^{2}})^{(1-D_m)/2} \cdot \\
	& \;\; \;\; \cdot \sin[(D_m-1)\arctan(Q\xi)], \nonumber
\end{align}
\end{subequations}
where $ \Delta \rho $ (=$7.5\cdot 10^{10} \, \text{cm}^{-2}$ for carbon-air) is the scattering length density contrast, $ V_p $ the volume of a primary pore building block,  $ \ell $ the associated length, $ P(Q) $ the form factor, characteristic of the surface fractal morphology of the pores,  $S(Q) $  the structure factor corresponding to the mass fractal structure 
and $ \Gamma $ the gamma function. Besides the fractal dimensions $ D_m $ and $ D_s $, two characteristic lengths are introduced: an upper cut-off length $\xi$ and  the length of the primary pore-building block $\ell$, which  marks the cross-over between mass and surface fractal scattering.\\
In order to fit all experimental results with this model in the most reliable way we  adopted the following strategy : (1) the values of $D_s$ were fixed to those of Table \ref{tab:power}, derived from SANS; (2) the values of $D_m$ and $\xi$ were derived from the scattering patterns of the Fourier transformed images. These are given in Table \ref{tab:power}, which shows that $ D_m\approx D_s $. For the sake of simplicity we assumed $ D_m=D_s $ in the following. (3) With  $D_m$, $D_s$ and $\xi$ fixed, $ \ell $ was determined by fitting the SESANS data, that fill the gap between SANS and imaging and probe the cross-over between mass and surface fractal scattering. For this purpose the Hankel transform was performed numerically on Eq. \ref{eq:SANS} (combined with Eq. \ref{eq:fractalmodel}) through Eq. \ref{eq:Henkeltrans}; then the data were fitted using Eq. \ref{eq:SESANS} with only two floating parameters: length $ \ell $ and the prefactor $\mathcal{B}$, and the resulting values are given in Table \ref{tab:param}. 
The fits are illustrated by the solid lines in Fig. \ref{fig:SESANS} and describe excellently the experimental findings.\\  
\begin{table}
	\caption{\label{tab:param} Parameters for the fractal model assuming $ D_s =D_m $. }
	\resizebox{\textwidth}{!}{
		\begin{tabular}{ccccccccc} \hline \hline
			& SANS & Imaging&\multicolumn{2}{c}{SESANS}&\multicolumn{3}{c}{Global}\\
			Sample&$ D_s =D_m $&$ \xi $ ($ \mu $m)&$ \ell $ ($ \mu $m)&$ \mathcal{B}$ (cm$ ^{-1}) $&$S(Q=0)$&$\mathcal{B}$ (cm$ ^{-1}) $&$\mathcal{B}$ (cm$ ^{-1})$\\ 
			&\textit{(fixed)}&\textit{(fitted)}&\textit{(fitted)}&\textit{(fitted)}&\textit{(calculated)}&\textit{(fitted)}&\textit{(calculated)}\\ \hline
			PGA1 & 2.55 & 18.9 (4) & 2.27 (10) & $ 6.6(1)\times 10^{10} $ & 134(15) & $ 7.56(5)\times 10^{10} $ & $ 6.3(9)\times 10^{10} $\\
			PGA2 & 2.57 & 17.0 (5) & 2.64 (25) & $ 8.4(1)\times 10^{10} $ & 74(5)   & $ 1.87(1)\times 10^{11} $ & $ 9.8(5)\times 10^{10} $\\
			RID  & 2.55 & 18.7 (4) & 1.87 (6)  & $ 3.0(1)\times 10^{10} $ & 214(17) & $ 4.64(4)\times 10^{10} $ & $ 4.2(5)\times 10^{10} $\\ \hline \hline
		\end{tabular}}
\end{table}
The final step is to combine  SANS, SESANS and imaging, which is done in Fig. \ref{fig:Global}. In this figure the lines correspond to the best fit of  Eq. \ref{eq:fractalmodel} with the parameters of Table \ref{tab:param}. The $Q$ values for which $Q\xi=1$ and $Q\ell=1$ are also indicated.  We stress that the model provides excellent quantitative description of the experimental findings. Indeed the prefactors $\mathcal{B}$ given in Table \ref{tab:param}, which are deduced from fitting all data (global)  are in excellent agreement with the calculated values from the fitted values of $\ell$ through Eq. \ref{eq:prefactor}. Therefore, a consistent picture of the fractal microstructure of graphite is obtained over an extraordinary large range of length scales of 6 orders of magnitude ($ \sim $0.6 nm $ \leq 2\pi/Q\leq$ 0.6 mm).
Table \ref{tab:param} shows that the fractal dimensions and cut-off lengths $ \xi $ are almost the same for all samples, which might be because both PGA and RID are made out of petroleum coke \cite{Martin1977JNM}. A trend is only seen in the values of $\ell$, which is the parameter that reflects the particularities of the microstructure and also the anisotropy of PGA2 (see supplement). It has been shown that the fractal anisotropy can be found in graphite at low densities (expanded graphite) and that in fact under densification such anisotropy is lost \cite{Balima2013,Balima2014}. \\
In the literature besides the micropores mentioned above, so-called Mrozowski cracks have been reported \cite{Wen2008,Jones2008,Karthik2012}, that result from the anisotropic thermal shrinkage of the layered graphite structure during cooling from graphitization temperatures. These have lengths from nm to $\mu$m that are in the range probed by   
SESANS and SANS. It is therefore tempting to attribute the surface fractal scattering to these cracks, as suggested recently \cite{Mileeva2013}, and $\ell$ to the cross-over between cracks and micropores.  
\\
In contrast to optical or electron microscopy, the scattering techniques used in this work cannot distinguish between cracks and pores. However, they provide a statistical average of the correlations over the (macroscopic) samples and reveal the most generic features of the structure. In this way it is possible to describe the complex and poly-disperse patterns of Fig. \ref{fig:Imaging1} with the scattering law of Eq. \ref{eq:fractalmodel} involving a restricted number of parameters, which enables quantitative comparisons between different samples and systems.  
\\
Besides the extraordinary large length scales over which fractality has been observed, the fractal dimensions found in this work are comparable to those expected for percolating clusters \cite{vanderMarck1997, Bradley1991}, for which heuristic arguments suggest that mass fractals should be bounded by their own natural fractal surface and $ D_m=D_s $ \cite{Bradley1991}. The porosity values of Table \ref{tab:power} indicate indeed a topology close to that of a percolating cluster \cite{vanderMarck1997}. \\
Similar fractal exponents have  been deduced for fracturing ranked surfaces in 3D \cite{Schrenk2012}, that could serve as a model for  Mrozowski cracks. Therefore the identity $ D_m=D_s\sim 2.5 $ is not coincidence but the consequence of the high degree of disorder, ramification and connectivity of the pore structure. Under neutron irradiation surface fractality disappears \cite{Mileeva2013}, and similar behaviour may be expected for the oxidised samples. The methodology developed in this work thus can be applied to further investigate the effect of irradiation damage and/or oxidation on the structural properties of graphite.\\
Fractal scattering has also been reported for carbon nanopores \cite{Pfeifer2002,Mileeva2012}, rocks \cite{Radlinski1999} or cement \cite{Allen2007}. The particularity of this work is in the extraordinarily broad length scale of six orders of magnitude over which  fractal scaling is quantitatively valid. The combination of several techniques, from imaging to  scattering and the  methods  can be applied to the investigation of other complex systems with a hierarchy of length scales such as biological materials, concrete and rocks, materials for CO2 sequestration, Li batteries, fuel cells or solar cells.
\section*{Acknowledgments}
We thank Annie Brulet, Chris Duif, Lucas J. van Vliet, Lambert van Eijck, Niels van Dijk, Ad van Well and Zhengcao Li for enlightening discussions and experimental support. We thank Paul Mummery for providing the PGA samples. Funding is acknowledged from the European Union Seventh Framework Programme [FP7/2007- 2013] under grant agreement Nr. 283883 and from the Royal Netherlands Academy of Arts and Sciences China Exchange Program  grant Nr. 12CDP011.
\section*{References}
\bibliography{myRef2}

\begin{thebibliography}{39}
\providecommand{\natexlab}[1]{#1}
\providecommand{\url}[1]{\texttt{#1}}
\providecommand{\href}[2]{#2}
\providecommand{\path}[1]{#1}
\providecommand{\eprint}[1]{\href{http://arxiv.org/abs/#1}{\path{#1}}}
\providecommand{\DOIprefix}{doi:}
\providecommand{\ArXivprefix}{arXiv:}
\providecommand{\URLprefix}{URL: }
\providecommand{\Pubmedprefix}{pmid:}
\providecommand{\doi}[1]{\href{http://dx.doi.org/#1}{\path{#1}}}
\providecommand{\Pubmed}[1]{\href{pmid:#1}{\path{#1}}}
\providecommand{\BIBand}{and}
\providecommand{\bibinfo}[2]{#2}
\ifx\xfnm\undefined \def\xfnm[#1]{\unskip,\space#1}\fi
\bibitem[{Haag(2005)}]{Haag2005}
\bibinfo{author}{Haag\xfnm[ G.]}.
\newblock \bibinfo{type}{Tech. Rep.} \bibinfo{number}{J{\"u}lich-4183};
  \bibinfo{year}{2005}.
\bibitem[{Zhou et~al.(2014)Zhou, Bouwman, Schut and Pappas}]{Zhou2014}
\bibinfo{author}{Zhou\xfnm[ Z.]}, \bibinfo{author}{Bouwman\xfnm[ W.G.]},
  \bibinfo{author}{Schut\xfnm[ H.]}, \bibinfo{author}{Pappas\xfnm[ C.]}.
\newblock \bibinfo{title}{Interpretation of x-ray diffraction patterns of
  (nuclear) graphite}.
\newblock \bibinfo{journal}{Carbon}
  \bibinfo{year}{2014};\bibinfo{volume}{69}:\bibinfo{pages}{17--24}.
\bibitem[{Wen et~al.(2008)Wen, Marrow and Marsden}]{Wen2008}
\bibinfo{author}{Wen\xfnm[ K.Y.]}, \bibinfo{author}{Marrow\xfnm[ T.J.]},
  \bibinfo{author}{Marsden\xfnm[ B.J.]}.
\newblock \bibinfo{title}{The microstructure of nuclear graphite binders}.
\newblock \bibinfo{journal}{Carbon}
  \bibinfo{year}{2008};\bibinfo{volume}{46}(\bibinfo{number}{1}):\bibinfo{pages}{62--71}.
\bibitem[{Jones et~al.(2008)Jones, Hall, Joyce, Hodgkins, Wen, Marrow
  et~al.}]{Jones2008}
\bibinfo{author}{Jones\xfnm[ A.N.]}, \bibinfo{author}{Hall\xfnm[ G.N.]},
  \bibinfo{author}{Joyce\xfnm[ M.]}, \bibinfo{author}{Hodgkins\xfnm[ A.]},
  \bibinfo{author}{Wen\xfnm[ K.]}, \bibinfo{author}{Marrow\xfnm[ T.J.]}, et~al.
\newblock \bibinfo{title}{Microstructural characterisation of nuclear grade
  graphite}.
\newblock \bibinfo{journal}{J Nucl Mater}
  \bibinfo{year}{2008};\bibinfo{volume}{381}(\bibinfo{number}{1-2}):\bibinfo{pages}{152--157}.
\bibitem[{Kane et~al.(2011)Kane, Karthik, Butt, Windes and Ubic}]{Kane2011}
\bibinfo{author}{Kane\xfnm[ J.]}, \bibinfo{author}{Karthik\xfnm[ C.]},
  \bibinfo{author}{Butt\xfnm[ D.P.]}, \bibinfo{author}{Windes\xfnm[ W.E.]},
  \bibinfo{author}{Ubic\xfnm[ R.]}.
\newblock \bibinfo{title}{Microstructural characterization and pore structure
  analysis of nuclear graphite}.
\newblock \bibinfo{journal}{J Nucl Mater}
  \bibinfo{year}{2011};\bibinfo{volume}{415}(\bibinfo{number}{2}):\bibinfo{pages}{189--197}.
\bibitem[{Karthik et~al.(2012)Karthik, Kane, Butt, Windes and
  Ubic}]{Karthik2012}
\bibinfo{author}{Karthik\xfnm[ C.]}, \bibinfo{author}{Kane\xfnm[ J.]},
  \bibinfo{author}{Butt\xfnm[ D.P.]}, \bibinfo{author}{Windes\xfnm[ W.E.]},
  \bibinfo{author}{Ubic\xfnm[ R.]}.
\newblock \bibinfo{title}{Microstructural characterization of next generation
  nuclear graphites}.
\newblock \bibinfo{journal}{Microsc Microanal}
  \bibinfo{year}{2012};\bibinfo{volume}{18}(\bibinfo{number}{2}):\bibinfo{pages}{272--278}.
\bibitem[{Hacker et~al.(2000)Hacker, Neighbour and McEnaney}]{Hacker2000}
\bibinfo{author}{Hacker\xfnm[ P.J.]}, \bibinfo{author}{Neighbour\xfnm[ G.B.]},
  \bibinfo{author}{McEnaney\xfnm[ B.]}.
\newblock \bibinfo{title}{The coefficient of thermal expansion of nuclear
  graphite with increasing thermal oxidation}.
\newblock \bibinfo{journal}{J Phys D: Appl Phys}
  \bibinfo{year}{2000};\bibinfo{volume}{33}(\bibinfo{number}{8}):\bibinfo{pages}{991--998}.
\bibitem[{Hoinkis(1996)}]{thrower1996chemistry}
\bibinfo{author}{Hoinkis\xfnm[ E.]}.
\newblock \bibinfo{title}{Small angle scattering of neutrons and x-rays from
  carbons and graphites}.
\newblock In: \bibinfo{editor}{Thrower\xfnm[ P.A.]}, editor.
  \bibinfo{booktitle}{Chemistry \& Physics of Carbon};
  vol.~\bibinfo{volume}{25}. \bibinfo{publisher}{CRC Press};
  \bibinfo{year}{1996}, p.~\bibinfo{pages}{71}.
\bibitem[{Martin and Henson(1964)}]{Martin1964}
\bibinfo{author}{Martin\xfnm[ D.G.]}, \bibinfo{author}{Henson\xfnm[ R.W.]}.
\newblock \bibinfo{title}{Scattering of long wavelength neutrons by defects in
  neutron-irradiated graphite}.
\newblock \bibinfo{journal}{Phil Mag}
  \bibinfo{year}{1964};\bibinfo{volume}{9}(\bibinfo{number}{100}):\bibinfo{pages}{659--672}.
\bibitem[{Martin and Caisley(1977{\natexlab{a}})}]{Martin1977Car}
\bibinfo{author}{Martin\xfnm[ D.G.]}, \bibinfo{author}{Caisley\xfnm[ J.]}.
\newblock \bibinfo{title}{Some studies of effect of irradiation on neutron
  small-angle scattering from graphite}.
\newblock \bibinfo{journal}{Carbon}
  \bibinfo{year}{1977}{\natexlab{a}};\bibinfo{volume}{15}(\bibinfo{number}{4}):\bibinfo{pages}{251--255}.
\bibitem[{Martin and Caisley(1977{\natexlab{b}})}]{Martin1977JNM}
\bibinfo{author}{Martin\xfnm[ D.G.]}, \bibinfo{author}{Caisley\xfnm[ J.]}.
\newblock \bibinfo{title}{Inference of coke source of nuclear graphites from
  neutron small-angle scattering measurements}.
\newblock \bibinfo{journal}{J Nucl Mater}
  \bibinfo{year}{1977}{\natexlab{b}};\bibinfo{volume}{67}(\bibinfo{number}{3}):\bibinfo{pages}{318--319}.
\bibitem[{Martin and Caisley(1978)}]{Martin1978}
\bibinfo{author}{Martin\xfnm[ D.G.]}, \bibinfo{author}{Caisley\xfnm[ J.]}.
\newblock \bibinfo{title}{Some studies of effect of thermal and radiolytic
  oxidation on neutron small-angle scattering from nuclear graphites}.
\newblock \bibinfo{journal}{Carbon}
  \bibinfo{year}{1978};\bibinfo{volume}{16}(\bibinfo{number}{3}):\bibinfo{pages}{199--203}.
\bibitem[{Mileeva et~al.(2013)Mileeva, Ross and King}]{Mileeva2013}
\bibinfo{author}{Mileeva\xfnm[ Z.]}, \bibinfo{author}{Ross\xfnm[ D.K.]},
  \bibinfo{author}{King\xfnm[ S.M.]}.
\newblock \bibinfo{title}{A study of the porosity of nuclear graphite using
  small-angle neutron scattering}.
\newblock \bibinfo{journal}{Carbon}
  \bibinfo{year}{2013};\bibinfo{volume}{64}:\bibinfo{pages}{20--26}.
\bibitem[{Bradley et~al.(1991)Bradley, Strenski and Debierre}]{Bradley1991}
\bibinfo{author}{Bradley\xfnm[ R.M.]}, \bibinfo{author}{Strenski\xfnm[ P.N.]},
  \bibinfo{author}{Debierre\xfnm[ J.M.]}.
\newblock \bibinfo{title}{Surfaces of percolation clusters in three
  dimensions}.
\newblock \bibinfo{journal}{Phys Rev B}
  \bibinfo{year}{1991};\bibinfo{volume}{44}(\bibinfo{number}{1}):\bibinfo{pages}{76--84}.
\bibitem[{Saleur and Duplantier(1987)}]{Saleur1987}
\bibinfo{author}{Saleur\xfnm[ H.]}, \bibinfo{author}{Duplantier\xfnm[ B.]}.
\newblock \bibinfo{title}{Exact determination of the percolation hull exponent
  in two dimensions}.
\newblock \bibinfo{journal}{Phys Rev Lett}
  \bibinfo{year}{1987};\bibinfo{volume}{58}(\bibinfo{number}{22}):\bibinfo{pages}{2325--2328}.
\bibitem[{Schrenk et~al.(2012)Schrenk, Araujo, Andrade and
  Herrmann}]{Schrenk2012}
\bibinfo{author}{Schrenk\xfnm[ K.J.]}, \bibinfo{author}{Araujo\xfnm[ N.A.M.]},
  \bibinfo{author}{Andrade\xfnm[ J.S.]}, \bibinfo{author}{Herrmann\xfnm[
  H.J.]}.
\newblock \bibinfo{title}{Fracturing ranked surfaces}.
\newblock \bibinfo{journal}{Scientific Reports}
  \bibinfo{year}{2012};\bibinfo{volume}{2}.
\bibitem[{Marsden and Wickham(2006)}]{2006IAEAcharacterization}
\bibinfo{author}{Marsden\xfnm[ B.]}, \bibinfo{author}{Wickham\xfnm[ A.J.]}.
\newblock \bibinfo{title}{Characterization, treatment and conditioning of
  radioactive graphite from decommissioning of nuclear reactors}.
\newblock \bibinfo{year}{2006}.
\bibitem[{Balima et~al.(2013)Balima, Pischedda, Le~Floch, Brulet, Lindner,
  Duclaux et~al.}]{Balima2013}
\bibinfo{author}{Balima\xfnm[ F.]}, \bibinfo{author}{Pischedda\xfnm[ V.]},
  \bibinfo{author}{Le~Floch\xfnm[ S.]}, \bibinfo{author}{Brulet\xfnm[ A.]},
  \bibinfo{author}{Lindner\xfnm[ P.]}, \bibinfo{author}{Duclaux\xfnm[ L.]},
  et~al.
\newblock \bibinfo{title}{An in situ small angle neutron scattering study of
  expanded graphite under a uniaxial stress}.
\newblock \bibinfo{journal}{Carbon}
  \bibinfo{year}{2013};\bibinfo{volume}{57}:\bibinfo{pages}{460--469}.
\bibitem[{Balima et~al.(2014)Balima, Le~Floch, San-Miguel, Lindner, Brulet,
  Duclaux et~al.}]{Balima2014}
\bibinfo{author}{Balima\xfnm[ F.]}, \bibinfo{author}{Le~Floch\xfnm[ S.]},
  \bibinfo{author}{San-Miguel\xfnm[ A.]}, \bibinfo{author}{Lindner\xfnm[ P.]},
  \bibinfo{author}{Brulet\xfnm[ A.]}, \bibinfo{author}{Duclaux\xfnm[ L.]},
  et~al.
\newblock \bibinfo{title}{Shear effects on expanded graphite under uniaxial
  pressure: An in situ small angle neutron scattering study}.
\newblock \bibinfo{journal}{Carbon}
  \bibinfo{year}{2014};\bibinfo{volume}{74}:\bibinfo{pages}{54--62}.
\bibitem[{Willis and Carlile(2009)}]{Willis2009book}
\bibinfo{author}{Willis\xfnm[ B.T.M.]}, \bibinfo{author}{Carlile\xfnm[ C.J.]}.
\newblock \bibinfo{title}{Experimental neutron scattering}.
\newblock \bibinfo{publisher}{Oxford Univ. Press}; \bibinfo{year}{2009}.
\bibitem[{Desert et~al.(2007)Desert, Thevenot, Oberdisse and
  Brulet}]{Desert2007}
\bibinfo{author}{Desert\xfnm[ S.]}, \bibinfo{author}{Thevenot\xfnm[ V.]},
  \bibinfo{author}{Oberdisse\xfnm[ J.]}, \bibinfo{author}{Brulet\xfnm[ A.]}.
\newblock \bibinfo{title}{The new very-small-angle neutron scattering
  spectrometer at laboratoire leon brillouin}.
\newblock \bibinfo{journal}{J Appl Crystallogr}
  \bibinfo{year}{2007};\bibinfo{volume}{40}:\bibinfo{pages}{S471--S473}.
\bibitem[{Rekveldt et~al.(2005)Rekveldt, Plomp, Bouwman, Kraan, Grigoriev and
  Blaauw}]{Rekveldt2005}
\bibinfo{author}{Rekveldt\xfnm[ M.T.]}, \bibinfo{author}{Plomp\xfnm[ J.]},
  \bibinfo{author}{Bouwman\xfnm[ W.G.]}, \bibinfo{author}{Kraan\xfnm[ W.H.]},
  \bibinfo{author}{Grigoriev\xfnm[ S.]}, \bibinfo{author}{Blaauw\xfnm[ M.]}.
\newblock \bibinfo{title}{Spin-echo small angle neutron scattering in delft}.
\newblock \bibinfo{journal}{Rev Sci Instrum}
  \bibinfo{year}{2005};\bibinfo{volume}{76}(\bibinfo{number}{3}).
\bibitem[{Andersson et~al.(2008)Andersson, van Heijkamp, de~Schepper and
  Bouwman}]{Anderson2008}
\bibinfo{author}{Andersson\xfnm[ R.]}, \bibinfo{author}{van Heijkamp\xfnm[
  L.F.]}, \bibinfo{author}{de~Schepper\xfnm[ I.M.]},
  \bibinfo{author}{Bouwman\xfnm[ W.G.]}.
\newblock \bibinfo{title}{Analysis of spin-echo small-angle neutron scattering
  measurements}.
\newblock \bibinfo{journal}{J Appl Crystallogr}
  \bibinfo{year}{2008};\bibinfo{volume}{41}:\bibinfo{pages}{868--885}.
\bibitem[{Kaestner et~al.(2011)Kaestner, Hartmann, Kuhne, Frei, Grunzweig,
  Josic et~al.}]{Kaestner2011}
\bibinfo{author}{Kaestner\xfnm[ A.P.]}, \bibinfo{author}{Hartmann\xfnm[ S.]},
  \bibinfo{author}{Kuhne\xfnm[ G.]}, \bibinfo{author}{Frei\xfnm[ G.]},
  \bibinfo{author}{Grunzweig\xfnm[ C.]}, \bibinfo{author}{Josic\xfnm[ L.]},
  et~al.
\newblock \bibinfo{title}{The icon beamline - a facility for cold neutron
  imaging at sinq}.
\newblock \bibinfo{journal}{NIMA}
  \bibinfo{year}{2011};\bibinfo{volume}{659}(\bibinfo{number}{1}):\bibinfo{pages}{387--393}.
\bibitem[{Schelten and Schmatz(1980)}]{Schelten1980}
\bibinfo{author}{Schelten\xfnm[ J.]}, \bibinfo{author}{Schmatz\xfnm[ W.]}.
\newblock \bibinfo{title}{Multiple-scattering treatment for small-angle
  scattering problems}.
\newblock \bibinfo{journal}{J Appl Crystallogr}
  \bibinfo{year}{1980};\bibinfo{volume}{13}:\bibinfo{pages}{385--390}.
\bibitem[{Radlinski et~al.(1999)Radlinski, Radlinska, Agamalian, Wignall,
  Lindner and Randl}]{Radlinski1999}
\bibinfo{author}{Radlinski\xfnm[ A.P.]}, \bibinfo{author}{Radlinska\xfnm[
  E.Z.]}, \bibinfo{author}{Agamalian\xfnm[ M.]}, \bibinfo{author}{Wignall\xfnm[
  G.D.]}, \bibinfo{author}{Lindner\xfnm[ P.]}, \bibinfo{author}{Randl\xfnm[
  O.G.]}.
\newblock \bibinfo{title}{Fractal geometry of rocks}.
\newblock \bibinfo{journal}{Phys Rev Lett}
  \bibinfo{year}{1999};\bibinfo{volume}{82}(\bibinfo{number}{15}):\bibinfo{pages}{3078--3081}.
\bibitem[{Rekveldt et~al.(2003)Rekveldt, Bouwman, Kraan, Uca, Grigoriev,
  Habicht et~al.}]{Rekveldt2003}
\bibinfo{author}{Rekveldt\xfnm[ M.T.]}, \bibinfo{author}{Bouwman\xfnm[ W.G.]},
  \bibinfo{author}{Kraan\xfnm[ W.H.]}, \bibinfo{author}{Uca\xfnm[ O.]},
  \bibinfo{author}{Grigoriev\xfnm[ S.V.]}, \bibinfo{author}{Habicht\xfnm[ K.]},
  et~al.
\newblock \bibinfo{title}{Elastic neutron scattering measurements using larmor
  precession of polarized neutrons}.
\newblock \bibinfo{journal}{Neutron Spin Echo Spectroscopy: Basics, Trends and
  Applications}
  \bibinfo{year}{2003};\bibinfo{volume}{601}:\bibinfo{pages}{87--99}.
\bibitem[{Mildner and Hall(1986)}]{Mildner1986}
\bibinfo{author}{Mildner\xfnm[ D.F.R.]}, \bibinfo{author}{Hall\xfnm[ P.L.]}.
\newblock \bibinfo{title}{Small-angle scattering from porous solids with
  fractal geometry}.
\newblock \bibinfo{journal}{J Phys D: Appl Phys}
  \bibinfo{year}{1986};\bibinfo{volume}{19}(\bibinfo{number}{8}):\bibinfo{pages}{1535--1545}.
\bibitem[{Schmidt(1991)}]{Schmidt1991}
\bibinfo{author}{Schmidt\xfnm[ P.W.]}.
\newblock \bibinfo{title}{Small-angle scattering studies of disordered, porous
  and fractal systems}.
\newblock \bibinfo{journal}{J Appl Crystallogr}
  \bibinfo{year}{1991};\bibinfo{volume}{24}:\bibinfo{pages}{414--435}.
\bibitem[{Bale and Schmidt(1984)}]{Bale1984}
\bibinfo{author}{Bale\xfnm[ H.D.]}, \bibinfo{author}{Schmidt\xfnm[ P.W.]}.
\newblock \bibinfo{title}{Small-angle x-ray-scattering investigation of
  submicroscopic porosity with fractal properties}.
\newblock \bibinfo{journal}{Phys Rev Lett}
  \bibinfo{year}{1984};\bibinfo{volume}{53}(\bibinfo{number}{6}):\bibinfo{pages}{596--599}.
\bibitem[{Teixeira(1988)}]{Teixeira1988}
\bibinfo{author}{Teixeira\xfnm[ J.]}.
\newblock \bibinfo{title}{Small-angle scattering by fractal systems}.
\newblock \bibinfo{journal}{J Appl Crystallogr}
  \bibinfo{year}{1988};\bibinfo{volume}{21}:\bibinfo{pages}{781--785}.
\bibitem[{Mandelbrot(1983)}]{Madelbrot1983}
\bibinfo{author}{Mandelbrot\xfnm[ B.B.]}.
\newblock \bibinfo{title}{The fractal geometry of nature}.
\newblock \bibinfo{address}{New York}: \bibinfo{publisher}{Freeman};
  \bibinfo{year}{1983}.
\bibitem[{Pentland(1984)}]{Pentland1984}
\bibinfo{author}{Pentland\xfnm[ A.P.]}.
\newblock \bibinfo{title}{Fractal-based description of natural scenes}.
\newblock \bibinfo{journal}{IEEE Transactions on Pattern Analysis and Machine
  Intelligence}
  \bibinfo{year}{1984};\bibinfo{volume}{6}(\bibinfo{number}{6}):\bibinfo{pages}{661--674}.
\bibitem[{Beech(1992)}]{Beech1992}
\bibinfo{author}{Beech\xfnm[ M.]}.
\newblock \bibinfo{title}{The projection of fractal objects}.
\newblock \bibinfo{journal}{Astrophys Space Sci}
  \bibinfo{year}{1992};\bibinfo{volume}{192}(\bibinfo{number}{1}):\bibinfo{pages}{103--111}.
\bibitem[{Wang et~al.(2013)Wang, Anovitz, Burg, Cole, Allard, Jackson
  et~al.}]{Wang2013}
\bibinfo{author}{Wang\xfnm[ H.W.]}, \bibinfo{author}{Anovitz\xfnm[ L.M.]},
  \bibinfo{author}{Burg\xfnm[ A.]}, \bibinfo{author}{Cole\xfnm[ D.R.]},
  \bibinfo{author}{Allard\xfnm[ L.F.]}, \bibinfo{author}{Jackson\xfnm[ A.J.]},
  et~al.
\newblock \bibinfo{title}{Multi-scale characterization of pore evolution in a
  combustion metamorphic complex, hatrurim basin, israel: Combining (ultra)
  small-angle neutron scattering and image analysis}.
\newblock \bibinfo{journal}{Geochim Cosmochim Acta}
  \bibinfo{year}{2013};\bibinfo{volume}{121}:\bibinfo{pages}{339--362}.
\bibitem[{van~der Marck(1997)}]{vanderMarck1997}
\bibinfo{author}{van~der Marck\xfnm[ S.C.]}.
\newblock \bibinfo{title}{Percolation thresholds and universal formulas}.
\newblock \bibinfo{journal}{Phys Rev E}
  \bibinfo{year}{1997};\bibinfo{volume}{55}(\bibinfo{number}{2}):\bibinfo{pages}{1514--1517}.
\bibitem[{Pfeifer et~al.(2002)Pfeifer, Ehrburger-Dolle, Rieker, Gonzalez,
  Hoffman, Molina-Sabio et~al.}]{Pfeifer2002}
\bibinfo{author}{Pfeifer\xfnm[ P.]}, \bibinfo{author}{Ehrburger-Dolle\xfnm[
  F.]}, \bibinfo{author}{Rieker\xfnm[ T.P.]}, \bibinfo{author}{Gonzalez\xfnm[
  M.T.]}, \bibinfo{author}{Hoffman\xfnm[ W.P.]},
  \bibinfo{author}{Molina-Sabio\xfnm[ M.]}, et~al.
\newblock \bibinfo{title}{Nearly space-filling fractal networks of carbon
  nanopores}.
\newblock \bibinfo{journal}{Phys Rev Lett}
  \bibinfo{year}{2002};\bibinfo{volume}{88}(\bibinfo{number}{11}).
\bibitem[{Mileeva et~al.(2012)Mileeva, Ross, Wilkinson, King, Ryan and
  Sharrock}]{Mileeva2012}
\bibinfo{author}{Mileeva\xfnm[ Z.]}, \bibinfo{author}{Ross\xfnm[ D.K.]},
  \bibinfo{author}{Wilkinson\xfnm[ D.]}, \bibinfo{author}{King\xfnm[ S.M.]},
  \bibinfo{author}{Ryan\xfnm[ T.A.]}, \bibinfo{author}{Sharrock\xfnm[ H.]}.
\newblock \bibinfo{title}{The use of small angle neutron scattering with
  contrast matching and variable adsorbate partial pressures in the study of
  porosity in activated carbons}.
\newblock \bibinfo{journal}{Carbon}
  \bibinfo{year}{2012};\bibinfo{volume}{50}(\bibinfo{number}{14}):\bibinfo{pages}{5062--5075}.
\bibitem[{Allen et~al.(2007)Allen, Thomas and Jennings}]{Allen2007}
\bibinfo{author}{Allen\xfnm[ A.J.]}, \bibinfo{author}{Thomas\xfnm[ J.J.]},
  \bibinfo{author}{Jennings\xfnm[ H.M.]}.
\newblock \bibinfo{title}{Composition and density of nanoscale
  calcium-silicate-hydrate in cement}.
\newblock \bibinfo{journal}{Nature Mat}
  \bibinfo{year}{2007};\bibinfo{volume}{6}(\bibinfo{number}{4}):\bibinfo{pages}{311--316}.

\end{thebibliography}
\end{document}